\documentclass[aps,prl,preprint,nopacs,amsmath,superscriptaddress]{revtex4}

\usepackage{epsf}
\usepackage{graphicx}
\usepackage{sidecap}
\usepackage{amsmath}

\usepackage{dcolumn}   
\usepackage{bm}        
\usepackage{amssymb}   

\begin{document}

\title{Observation of Quantum-Tunneling Modulated Spin Texture in Ultrathin Topological Insulator Bi$_2$Se$_3$ Films}

\author{Madhab~Neupane}
\affiliation {Joseph Henry Laboratory and Department of Physics,
Princeton University, Princeton, New Jersey 08544, USA}

\author{Anthony Richardella}
\affiliation{Department of Physics, The Pennsylvania State University, University Park, PA 16802}

\author{J. S\'anchez-Barriga}
\affiliation {Helmholtz-Zentrum Berlin f\"ur Materialien und Energie, Elektronenspeicherring BESSY II, Albert-Einstein-Strasse 15, D-12489 Berlin, Germany}

\author{SuYang~Xu}
\affiliation {Joseph Henry Laboratory and Department of Physics,
Princeton University, Princeton, New Jersey 08544, USA}

\author{Nasser~Alidoust}
\affiliation {Joseph Henry Laboratory and Department of Physics,
Princeton University, Princeton, New Jersey 08544, USA}

\author{Ilya~Belopolski}
\affiliation {Joseph Henry Laboratory and Department of Physics,
Princeton University, Princeton, New Jersey 08544, USA}

\author{Chang~Liu}
\affiliation {Joseph Henry Laboratory and Department of Physics,
Princeton University, Princeton, New Jersey 08544, USA}

\author{Guang~Bian}
\affiliation {Joseph Henry Laboratory and Department of Physics,
Princeton University, Princeton, New Jersey 08544, USA}

\author{ Duming Zhang}
\affiliation{Department of Physics, The Pennsylvania State University, University Park, PA 16802}

\author{Dmitry Marchenko}
\affiliation {Helmholtz-Zentrum Berlin f\"ur Materialien und Energie, Elektronenspeicherring BESSY II, Albert-Einstein-Strasse 15, D-12489 Berlin, Germany}

\affiliation {Physikalische und Theoretische Chemie, Freie Universit\"at Berlin, Takustra{\ss}e 3, 14195 Berlin, Germany}


\author{Andrei Varykhalov}
\affiliation{Helmholtz-Zentrum Berlin f\"ur Materialien und Energie, Elektronenspeicherring BESSY II, Albert-Einstein-Strasse 15, D-12489 Berlin, Germany}

\author{Oliver Rader}
\affiliation {Helmholtz-Zentrum Berlin f\"ur Materialien und Energie, Elektronenspeicherring BESSY II, Albert-Einstein-Strasse 15, D-12489 Berlin, Germany}

\author{Mats Leandersson}
\affiliation {MAX-lab, P.O. Box 118, S-22100 Lund, Sweden}

\author{Thiagarajan Balasubramanian}
\affiliation {MAX-lab, P.O. Box 118, S-22100 Lund, Sweden}

\author{Tay-Rong Chang}
\affiliation{Department of Physics, National Tsing Hua University, Hsinchu 30013, Taiwan}

\author{Horng-Tay Jeng}
\affiliation{Department of Physics, National Tsing Hua University, Hsinchu 30013, Taiwan}
\affiliation{Institute of Physics, Academia Sinica, Taipei 11529, Taiwan}

\author{Susmita~Basak}
\affiliation {Department of Physics, Northeastern University,
Boston, Massachusetts 02115, USA}

\author{Hsin~Lin}
\affiliation {Department of Physics, Northeastern University,
Boston, Massachusetts 02115, USA}

\author{Arun~Bansil}
\affiliation {Department of Physics, Northeastern University,
Boston, Massachusetts 02115, USA}

\author{Nitin Samarth}
\affiliation{Department of Physics, The Pennsylvania State University, University Park, PA 16802}

\author{M.~Zahid~Hasan}
\affiliation {Joseph Henry Laboratory and Department of Physics,
Princeton University, Princeton, New Jersey 08544, USA}
\affiliation {Princeton Center for Complex Materials, Princeton University, Princeton, New Jersey 08544, USA}

\date{\today}

\newpage

\begin{abstract}
\textbf{Understanding the spin-texture behavior of boundary modes in ultrathin topological insulator films is critically essential for the design and fabrication of functional nano-devices. Here by using spin-resolved photoemission spectroscopy with p-polarized light in topological insulator Bi$_2$Se$_3$ thin films, we report tunneling-dependent evolution of spin configuration in topological insulator thin films across the metal-to-insulator transition. We observe strongly binding energy- and wavevector-dependent spin polarization for the topological surface electrons in the ultra-thin gapped-Dirac-cone limit. The polarization decreases significantly with enhanced tunneling realized systematically in thin insulating films, whereas magnitude of the polarization saturates to the bulk limit faster at larger wavevectors in thicker metallic films. We present a theoretical model which captures this delicate relationship between quantum tunneling and Fermi surface spin polarization. Our high-resolution spin-based spectroscopic results suggest that the polarization current can be tuned to zero in thin insulating films forming the basis for a future spin-switch nano-device.}

\end{abstract}


\maketitle

A three-dimensional topological insulator (TI) is a non-trivial phase of matter which acts as an electrical insulator in the bulk but can conduct a spin-polarized current on the surface \cite{Moore Nature insight, Zahid, Zhang, Fu Liang PRB topological invariants, David Nature BiSb, Matthew Nature physics BiSe, YLChen, SCZhang, David Nature tunable, SCZhang1, TCI, QiKunXue, SmB6_Hasan, Sakamoto,Tjernberg, SbTe, Ando,Neupane, QKXue}. These topological surface states are characterized by a Dirac cone-like energy-momentum dispersion relation. The novel electronic structure of TIs can be manipulated to realize various novel quantum phenomena such as  spin-galvanic effects, dissipationless spin currents or neutral half-fermions for quantum information storage devices \cite{Galvanic effect, Qi Science Monopole, Essin PRL Magnetic, Yu Science QAH, Liang Fu PRL Superconductivity, Linder PRL Superconductivity}. The magnitude and wavevector dependence of the spin polarization of electrons and holes are among the most important  key ingredients in considerations for the design of any functional device. However, such developments have been limited due to the residual bulk conductance in currently available materials which overwhelms the surface contribution. Additionally, scattering from the extrinsic bulk states leads to the reduction of spin polarization of the surface states. One promising route to minimize bulk conductance and thus improve effective spin polarization is to work with ultrathin films where the surface to volume ratio is significantly enhanced \cite{Linder, HZLu} and surface current can potentially dominate. On the other hand, in this limit the desired spin polarization of the surface states is kinematically reduced near the metal-to-insulator transition in the ultrathin films where the spin behavior is not known to this date \cite{QiKunXue, Sakamoto,Tjernberg,QKXue,SbTe}.

To date, no systematic spin-sensitive spectroscopic experimental study has been reported in the ultrathin limit across the metal-to-insulator transition despite the direct relevance of spins in ultrathin film limits for nano-device fabrication as well as the  potential discovery of novel topological phenomena. Studying the spin polarization in the ultra-thin limit is further important to experimentally demonstrate the theoretically predicted tunable Berry's phase in TI thin films \cite{loca}. 
Systematic mapping of the surface spin texture in energy, momentum space and its thickness dependence is essential to understand and interpret many transport experiments on thin film TIs that are of core interest in the current TI research.
We report a systematic spin-resolved angle-resolved photoemission spectroscopy (SR-ARPES) and spin-integrated ARPES measurements on ultrathin Bi$_2$Se$_3$ thin films for the first time. Our measurements reveal that the spin polarization is large for larger wavevectors, and the polarization magnitude increases with reduced tunneling, and its magnitude saturates to the bulk limit at a faster rate at large electron momenta. 

We observe strongly binding energy- and wavevector-dependent spin polarization for the topological surface electrons in the ultra-thin gapped-
Dirac-cone limit, which experimentally shows that the Dirac gap opening and the thickness-dependent topological phase transition are a result of the quantum tunneling between the two oppositely spin-textured topological
surface states. These unique spin features of ultrathin films, evidently distinct from the 3D TI, open up new possibilities for devices not possible with bulk topological insulators.

\bigskip
\bigskip
\textbf{Results}
\newline
\textbf{Sample characterization}

Spectroscopic measurements were performed on large ultrathin Bi$_2$Se$_3$ films prepared by the  Molecular Beam Epitaxy (MBE) method on GaAs(111)A substrates (Fig. 1a). Each crystal layer of Bi$_2$Se$_3$ constitutes of five atomic  layer, namely Se-Bi-Se-Bi-Se, which is called quintpule layer (QL) with the thickness of approximately 1 nm \cite{Matthew Nature physics BiSe}. 
Our MBE films grow in a self-organized quintuple-layer by quintuple-layer mode, and high quality atomically smooth film can be obtained with the desired thickness (also see Supplementary Figures 1-7 and Supplementary Notes 1-3). 
A compositional layout of the  film used in our measurements is shown in Fig. 1(b).
To protect the surface from contamination, about 40 nm Se capping is used on the top of ultrathin Bi$_2$Se$_3$ films. To expose the surface, the films were  transferred into the ARPES chamber and heated to 250 $^{\circ}$C at pressures lower than 1$\times$10$^{-9}$ torr about an hour which blow off the Se capping layer.
Fig. 1(c) shows the ARPES core level spectroscopy measurement of the unltrathin film before and after decapping of Se layer. Before decapping only Se peaks are visible while both Se and Bi peaks are observed after the decapping process, which proves that the Se capping works well in ultrathin film TI system. Thin films are characterized  by atomic force microscopy (AFM) (see Fig. 1(d)) and show that the root mean square (rms) surface roughness on these films is in the order of $\sim$ 0.2 nm, which confirms the high quality of the films used in our measurements. The transport measurements of the Se capped ultrathin films result in carrier concentration, mobility and resistivity in the order of 1$\times$10$^{19} $cm$^{-3}$, 1270 cm$^{-2}$.V$^{-1}$.Sec$^{-1}$ and 0.30 mOhm$\cdot$cm, respectively (also see Supplementary Figure 1 and Supplementary Note 1).

\textbf{Geometry of the spectroscopic measurements}

We used p-polarized light for our ARPES measurements. The photons approach the sample surface with angle of incident ($\theta$) 45 degree with the sample normal  and our samples are aligned along $\bar{\Gamma} - \bar{\text{K}}$ momentum-space cut (Fig. 2a) for spin-ARPES measurements.
The surface wavevector dependent  spin polarization is obtained using a Mott-polarimeter (Fig. 2(a)), which measures two orthogonal spin components of a photoemitted electron \cite{Berntsen, Hugo}.
 In the polarimeter, a gold foil was used as a scattering source to generate an asymmetry of high energy photoelectrons into different divergent spin states (see \cite{Hugo, Berntsen}  for details). Each orthogonal spin-polarization component is selected by the orientation of a scattering plane defined by the incident beam direction of the photoelectron on the gold foil and the orientation of two electron detectors mounted on each side of the foil. For this experiment, the detector configuration was set in a way that the two spin components correspond to the in-plane and out-of-plane directions of the (111) plane of the sample.

 \bigskip
\bigskip
\textbf{Thickness and wavevector dependent spin polarization}

We present high-resolution spin-integrated ARPES data  and corresponding energy distribution curves (EDCs) along the high symmetry line $\bar{\Gamma} - \bar{\text{K}}$ for 1QL, 3QL, 4QL, 6QL and 7QL Bi$_2$Se$_3$ films in Figs. 2(c) and 2(d).
As long as the thickness of the film is comparable to the decay length of the surface states into the bulk, there is a spatial overlap between the top and bottom surface states resulting in an energy gap at the time-reversal invariant point ($\bar\Gamma$ point).
As expected theoretically \cite{Linder, HZLu}, the energy gap decreases and eventually vanishes for sufficiently thick films, corresponding to the transition from a 2D gapped system (insulator) to a 3D gapless system (metal) \cite{QiKunXue, Sakamoto}. In particular, the gapless dispersion relation observed in the 7QL film from ARPES measurement indicates that this thickness is above the quantum tunneling limit.
Based on experimental observations, we present an illustration of the spin configuration for 3QL (insulator) (see Supplementary Figures 2 and 3, and Supplementary Note 2) and 7QL (metal) ultrathin Bi$_2$Se$_3$ films in Fig. 2(b). It is important to note that
a wide range of electronic structures have been reported in ultrathin TI films grown by MBE depending on the nature of the substrates used \cite{QiKunXue, Sakamoto, Tjernberg}. Different substrates result in different potential jump from the substrate (the bottom surface of the ultra-thin film) to the vacuum (the top surface of the film). When the potential is large, the Dirac point energy of the bottom surface is offset with respect to that of the top surface. Such Dirac point energy offset is observed to cause sizeable Rashba-type splitting of surface states as reported in ultrathin Bi$_2$Se$_3$ films grown on
double-layer-graphene-terminated 6H-SiC (0001) substrate \cite{QiKunXue} and Bi-terminated Si(111)-(7$\times$7) \cite{Tjernberg} substrate by MBE, respectively. On the other hand, no observable Rashba-type splitting due to substrate potential is reported in the Bi$_2$Se$_3$ grown on Si(111)$\beta\sqrt{3}\times\sqrt{3}$-Bi substrate (in Ref. \cite{Sakamoto}), $\alpha$-Al$_2$O$_3$ (sapphire) (0001) substrate (in Ref. \cite{QKXue}) and our current work GaAs(111)A.


We investigate the degree of spin polarization of the films as a function of electron wavevector for films of various thickness. To illustrate the wavevector dependent spin polarization, 3QL film is chosen ( see Supplementary Figures 2 and 3, and Supplementary Note 2 for details).
The SR-ARPES data for the wavevectors $\sim$ 0.05 \AA$^{-1}$ and 0.1 \AA$^{-1}$  are presented in Fig. 3. For each wavevector selected we present a spin-resolved energy distribution curve (EDC), which shows the relative intensity of photoelectrons with up and down spin polarizations (Figs. 3(a) and 3(c)).
For each such plot we associate a net spin polarization with the surface state of a given wavevector (Figs. 3(b) and 3(d)) and for 3QL film about 25 $\%$ and 15 $\%$ are estimated at momenta $\sim$ 0.1 \AA$^{-1}$  and $\sim$ 0.05 \AA$^{-1}$, respectively. From these plots it is clearly observed that the net spin polarization decreases for smaller wavevectors (locations k $\sim$ 0.05 \AA$^{-1}$ in Fig. 3(d)).
The decrease can be understood as the presence of a tunneling gap in the ultrathin limit that effectively prevents the partner-switching behavior expected in the gapless topological surface states system \cite{HZLu}. The tunneling gap for ultrathin films can be seen in  the data (Fig. 2(c)).
The reduction of the spin polarization and the existence of a tunneling gap in the data suggest that the bottom surface's right-handed contribution to the spin polarization must increase to effectively cancel the top surface's left-handed helical spin texture upon approaching smaller momenta values near $\bar{\Gamma}$.

Fig. 3(a) shows the spin-resolved EDCs  for 3QL, 4QL, 6QL and 60QL films while the corresponding net polarization is shown in Fig. 3(b) at $k$ = 0.1 \AA$^{-1}$. Analogous measurements are shown in Figs. 3(c) and 3(d) at $k$ = 0.05 \AA$^{-1}$ for 1QL, 3QL, 4QL, 6QL and 60QL  films.
For thinner films, such as 3QL, there is a naturally larger contribution from the bottom surface (due to tunneling between top and bottom surfaces). This leads to a larger variation in spin polarization as a function of wavevector. For thicker films, this tunneling contribution decreases and the surface spin polarization becomes more uniform with varying wavevector. For instance, in the 60QL film, no measurable change in spin polarization is observed for the variation in wavevector magnitude. Our data suggest that the magnitude of polarization tends to reach the bulk limit faster at larger wavevectors.
 Experimental results are summarized in Fig. 4(a). Data further suggest that the relative contribution from the top surface systematically increases with film thickness.

\bigskip
\bigskip
\textbf{Model calculation}

First-principles theoretical modeling of the spin polarization behavior in thin films is presented in Figs. 4(b) and 4(c).
In the calculations, symmetric slabs are used to simulate the thickness of  films. While a gapless spin-polarized Dirac cone is seen on the surface of a semi-infinite crystal of the topological insulator Bi$_2$Se$_3$, a gap is found to  open at the Dirac point for thin films due to a finite tunneling amplitude between the two sides of the slab surface in our calculations. The tunneling amplitude increases as the thickness decreases, causing  the gap to increase and
spin polarization to decrease in the gap region. In the calculation we also consider the electron attenuation length ($\lambda$) due to the scattering processes since only electrons near the surface are able to reach the top of the surface and escape into the vacuum as in the measurement condition. Indeed, the spin polarization obtained by ARPES reflects the spin-texture of the states associated with top surface rather than the bottom surface.
The calculated spin expectation value for the electrons that can escape from the sample is  $\langle S \rangle_{\text{atom}} \times \exp ({-d_{\text{atom}}/\lambda}$), where $d_{\text{atom}}$ is the distance of an atom to the top surface, and the $\langle S \rangle_{\text{atom}}$ is the spin expectation value for each atom. The contribution from each atom is weighted by exp($-d_{\text{atom}}/\lambda$), which reduces the contribution from the bottom layer. Figs. 4(b) and 4(c) show the calculated results with $\lambda$ =
8\AA, which agree excellently with our experimental observation
(see Supplementary Methods for detailed related to calculations).

\bigskip
\bigskip
\textbf{Discussion}

It is important to note that the maximum spin polarization observed in the bulk limit is only about 40\% (Fig. 4(c)) whereas the original ideal theoretical limit is that of nearly 100\%  without considering any specific material system \cite{Kane}.
 In real TI materials, the strong spin-orbit interaction entangles the spin and orbital momenta of different atomic types, resulting in the reduction of spin polarization \cite{spin}. Specifically, the low-energy states in Bi$_2$Se$_3$ arise from $p$-orbitals of Bi (6$p^3$) and Se (4$p^4$), mostly $p_z$ levels of Bi and Se \cite{Matthew Nature physics BiSe, SCZhang}. The dominance of the $p_z$ orbitals in the topological surface states is further suggested by our circular dichroism (CD) measurements (see Supplementary Figure 4, Supplementary Note 2 and Supplementary Methods) . The spin-orbit coupling mixes spin and orbital angular momenta while preserving the total angular momentum \cite{Matthew Nature physics BiSe, SCZhang}. The hybridization of orbitals in Bi and Se together with the entanglement of their spins contribute to the reduction of net spin polarization in real materials.
Moreover, under the experimental geometry used in our measurement with $p-$polarized light (which is most sensitive to $p_z$ orbitals and most reflective of initial groundstate of the wavefunction), the penetration depth of the ARPES experiment (3-5 atomic layers maximum), the experimental observation of spin polarization is well agreed with recent theoretical calculations \cite{Park, Oliver}.

Most importantly, our systematic spin-spectroscopy results suggest that ultrathin films can serve as the basis for making qualitatively new devices, not possible with much-studied conventional 3D bulk topological insulators, despite the reduction of the polarization magnitude. The energy-momentum space spin texture revealed in our study provides critical knowledge to design and interpret devices based on ultrathin films. For spintronics applications, our results suggest that ultrathin topological insulator films can be used to fabricate new types of nano-devices due to the novel spin configurations and their systematic 
modulations possible in the ultrathin limit. One such potential application implied by our spectroscopic results is that of a polarization current switch. Spin spectroscopic results suggest that it should be possible to control the polarization magnitude by varying a gate voltage (Figs. 4(d) and 4(e)) across a high quality insulating thin film (see supplementary Figure 5 and Supplementary Note 3). Such electrostatic gating would be similar to the in-situ chemical gating by K-deposition and NO2 adsorption shown in Supplementary Figures 6 and 7, where the surface state structure does not change appreciably, while the Fermi energy is moved from regions of high polarization to low polarization (see Supplementary Note 3). This effect, which cannot be readily realized in the highly-polarized states of conventional 3D bulk TIs, allows ultrathin TI films to serve as the basis for functional nano-devices which can encode electrical signals in varying spin polarization magnitude or forms the physical basis for a spin switch, among many other new application possibilities suggested by our observations of the fundamental spin modulation behavior in ultrathin films.

\bigskip
\bigskip
\textbf{Methods}

\textbf{Sample growth and transport measurements}

Ultrathin Bi$_2$Se$_3$ films used for this study were synthesized by molecular beam epitaxy (MBE) on a GaAs(111)A substrate with a ZnSe buffer layer. Details of sample preparation are described elsewhere \cite{Suyang, DZhang}.
To protect the surface from oxidation, a thick Se capping layer was deposited on the Bi$_2$Se$_3$ thin film immediately after growth.  To achieve the clean Bi$_2$Se$_3$ surface required for photoemission measurements, thin films were heated up inside the ARPES chamber to 250 $^\circ$C under vacuum better than 1 $\times$ 10$^{-9}$ torr to remove the Se capping layer on top of ultrathin Bi$_2$Se$_3$.
Hall bars fabricated from the Se capped samples had carrier concentrations in the 1 to 2 $\times$10$^{19}$cm$^{-3}$ range.
Transport measurements were carried out on Hall bars of the 5 and 6QL thick films with the Se capping layer in place by standard photolithography and dry etching and were measured at 4.2 K.

\bigskip
\textbf{High-resolution ARPES measurements}

High-resolution spin-integrated angle-resolved photoemission spectroscopy (ARPES) measurements were performed with 29-64 eV photon energy on beamlines 10.0.1 and 12.0.1 at the Advanced Light Source (ALS) in Lawrence Berkeley National Laboratory (LBNL). Both endstations were equipped with a Scienta hemispherical electron analyser (see VG Scienta manufacturer website,
(http://www.vgscienta.com/), for instrument specifications).
The typical energy and momentum resolutions were
15 meV and 1\% of the surface Brillouin zone, respectively for spin-integrated measurements.

\bigskip
\textbf{Spin-resolved ARPES measurements}

Spin-resolved ARPES measurements were performed at the UE112-PGM1 beamline at Bessy II in Berlin, Germany, and the I3 beamline at Maxlab in Lund, Sweden, using classical Mott detectors and photon energies 55 to 60 eV and 8 to 20 eV, respectively.
The typical energy and momentum resolutions were
100 meV and 3\% of the surface Brillouin zone for spin-resolved measurements.
All the SR-spectra presented are measured in BESSY II unless it is specified.



\bigskip
\textbf{Theoretical calculations}

The first-principles calculations for spin polarizations of ultrathin films are based on the generalized
gradient approximation (GGA) \cite{Perdew} using the full-potential projected
augmented wave method \cite{Blochl, Bloch_1} as implemented in the VASP package \cite{Kress}. The 1QL, 3QL, 4QL, 6QL, 7QL, and 12QL slab models with a
vacuum thickness larger than 10 $\mathrm{\AA}$ are used in this work. The electronic structure calculations were performed over 11$\times$11$\times$1 Monkhorst-Pack k-mesh with the spin-orbit coupling included self-consistently.

\bigskip
\bigskip

\textit{Acknowledgements.}
Sample growth and ARPES characterization are supported by US DARPA (N66001-11-1-4110).
The work at Princeton University and Princeton-led synchrotron X-ray-based measurements and the related theory at Northeastern University are supported by the Office of Basic Energy Science, US Department of Energy (grants DE-FG-02-05ER462000, AC03-76SF00098 and DE-FG02-07ER46352).
M.Z.H. acknowledges visiting-scientist support from Lawrence
Berkeley National Laboratory and additional support from the
A. P. Sloan Foundation.
The spin-resolved and spin-integrated
photoemission measurements using synchrotron X-ray facilities are supported by
the Swedish Research Council, the Knut and Alice Wallenberg Foundation, the German Federal Ministry
of Education and Research, and the Basic Energy Sciences of the US Department of
Energy. Theoretical computations are supported by the US Department of Energy
(DE-FG02-07ER46352 and AC03-76SF00098) as well as the National Science Council and
Academia Sinica in Taiwan, and benefited from the allocation of supercomputer time at
NERSC and Northeastern University's Advanced Scientific Computation Center.
We also thank S-K. Mo and A. Fedorov for beamline assistance on spin-integrated photoemission
measurements (supported by DE-FG02-05ER46200) at Lawrence Berkeley National
Laboratory (The synchrotron facility is supported by the US DOE).

\bigskip

\textbf{Author contributions}
\newline
M.N., J.S.-B, and S.-Y.X. performed the experiments with assistance from N.A., I.B., C.L., G.B., D.M., A.V., O.R., M.L., T.B. and M.Z.H.;  A.R, D.M.Z and N. S. provided thin film MBE samples and performed sample characterization; M. N. and M. Z. H performed data analysis, figure planning and draft preparation.
T.R.C., H.T.J., S.B., H.L., and A.B. carried out calculations; M.Z.H. was responsible for the conception and the overall direction, planning and integration among different research units.

\textbf{Additional information}
\newline
Supplementary Information accompanies this paper is available at http://www.nature.com/naturecommunications.
\newline
Competing financial interests: The authors declare no competing financial interests.
\newline
\*Correspondence and requests for materials should be addressed to
\newline
M.Z.H. (Email: mzhasan@princeton.edu).

\newpage

\begin{figure*}
\includegraphics[width=17cm]{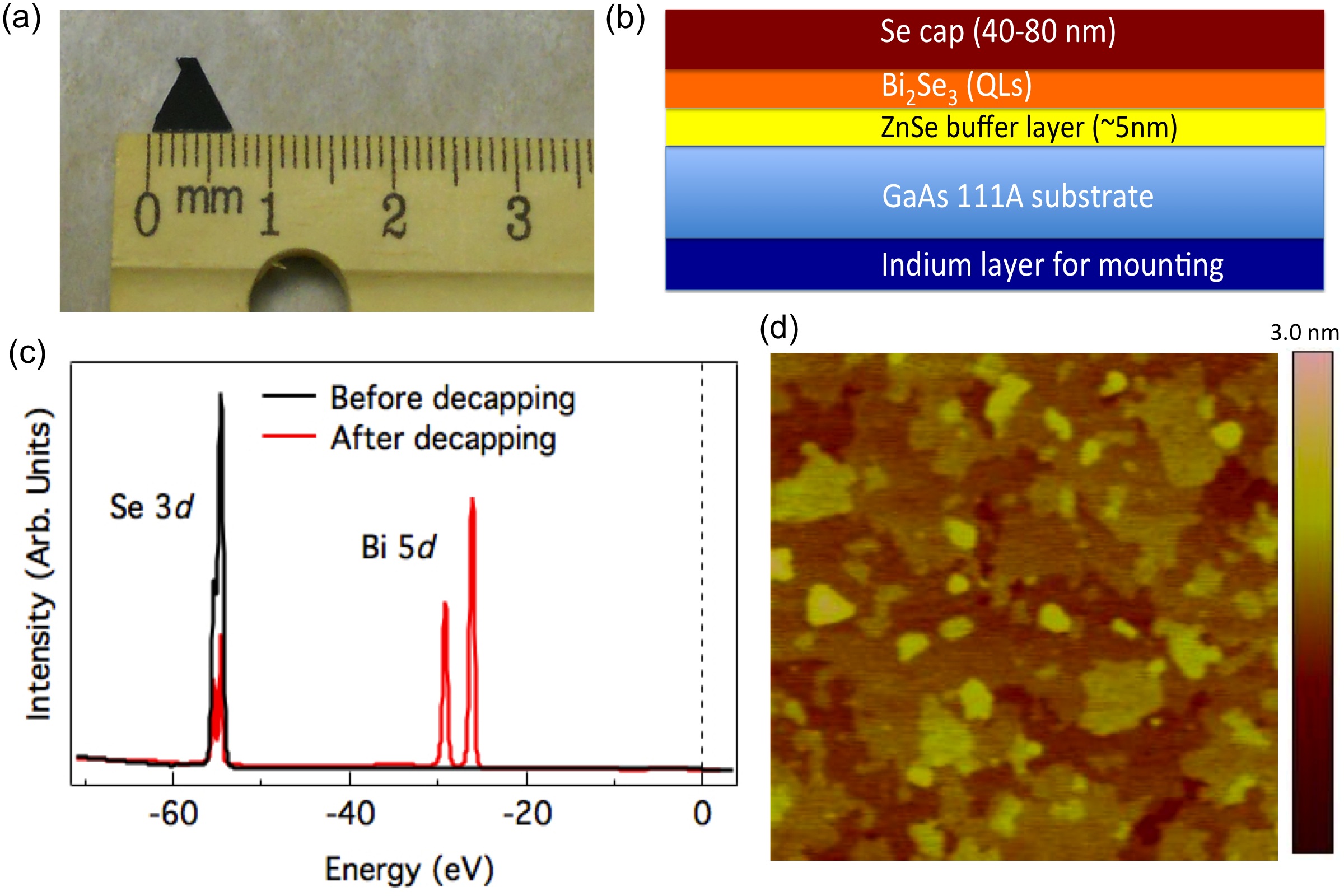}
\caption{\textbf{Characterization of MBE grown ultrathin films of Bi$_2$Se$_3$}.
(a) Photograph of a representative thin film sample used in SR-ARPES measurements.
(b) Sample layout of ultrathin Bi$_2$Se$_3$ film, grown by MBE.
(c) Core level spectroscopy measurements on ultrathin MBE film before and after the decapping procedure.
(d) AFM image of ultrathin Bi$_2$Se$_3$ film. The size of the image is 1 micron $\times$ 1 micron and height is indicated by the color bar on the right.}
\end{figure*}

\begin{figure*}
\centering
\includegraphics[width=16cm]{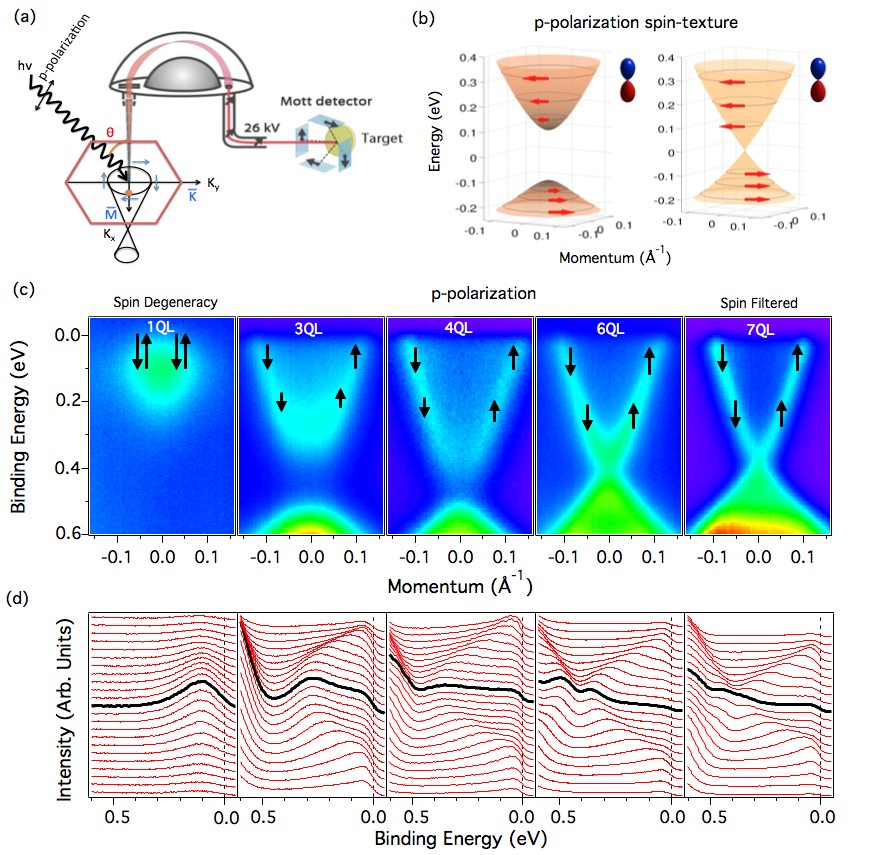}
\caption{\textbf{Spin-texture versus quantum tunneling in ultrathin Bi$_2$Se$_3$}.
(a) Experimental geometry used in our measurements.
(b) Visualization of the contrasting spin configurations in 3QL (insulator) and 7QL (metal)  thin films. The dumbbell signs indicate that the current experimental geometry mainly probes the $p_z$ orbitals of Bi and Se.
(c) High-resolution ARPES measurements on ultrathin films of Bi$_2$Se$_3$:  $E-k$ band dispersion images for 1QL, 3QL, 4QL, 6QL and 7QL of  Bi$_2$Se$_3$ films taken near the $\bar{\Gamma}$ point along $\bar{\Gamma}-\bar{\text{K}}$ high-symmetry direction. The spin configuration is noted on the plots. These spectra are measured with photon energy of 60 eV.
(d) The corresponding energy distribution curves (EDCs). The EDC through the $\bar{\Gamma}$ point (solid black curve) is highlighted.}
\end{figure*}

\begin{figure*}
\centering
\includegraphics[width=16cm]{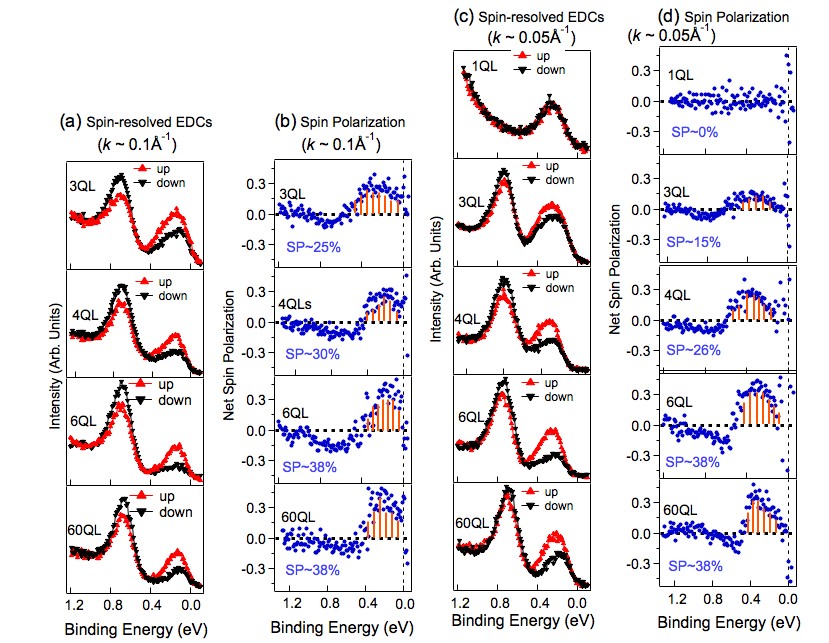}
\caption{\textbf{Thickness dependent quantum tunneling and evolution of spin configuration.}
(a) Spin-resolved EDCs and (b) net tangential spin polarizations for 3QL, 4QL, 6QL and 60QL ultrathin Bi$_2$Se$_3$ films at \textit{k} $\sim$ 0.1 \AA$^{-1}$. (c-d), same as (a) and (b) for 1QL, 3QL, 4QL, 6QL and 60QL ultrathin Bi$_2$Se$_3$ films at \textit{k} $\sim$ 0.05 \AA$^{-1}$.
The red (black) curves show tangentially up (down) spin-resolved EDCs. The magnitude of each net spin polarization is also noted in (b) and (d).
The vertical red lines in  (b) and (d) are guides to the eye, indicating a non-zero area under the spin polarization curve. SR-ARPES data were collected using photon energy of 60 eV.
We note that 1 quintuple layer (QL) is equivalent to $\sim$ 1 nm.}
\end{figure*}

\begin{figure*}
\includegraphics[width=16cm]{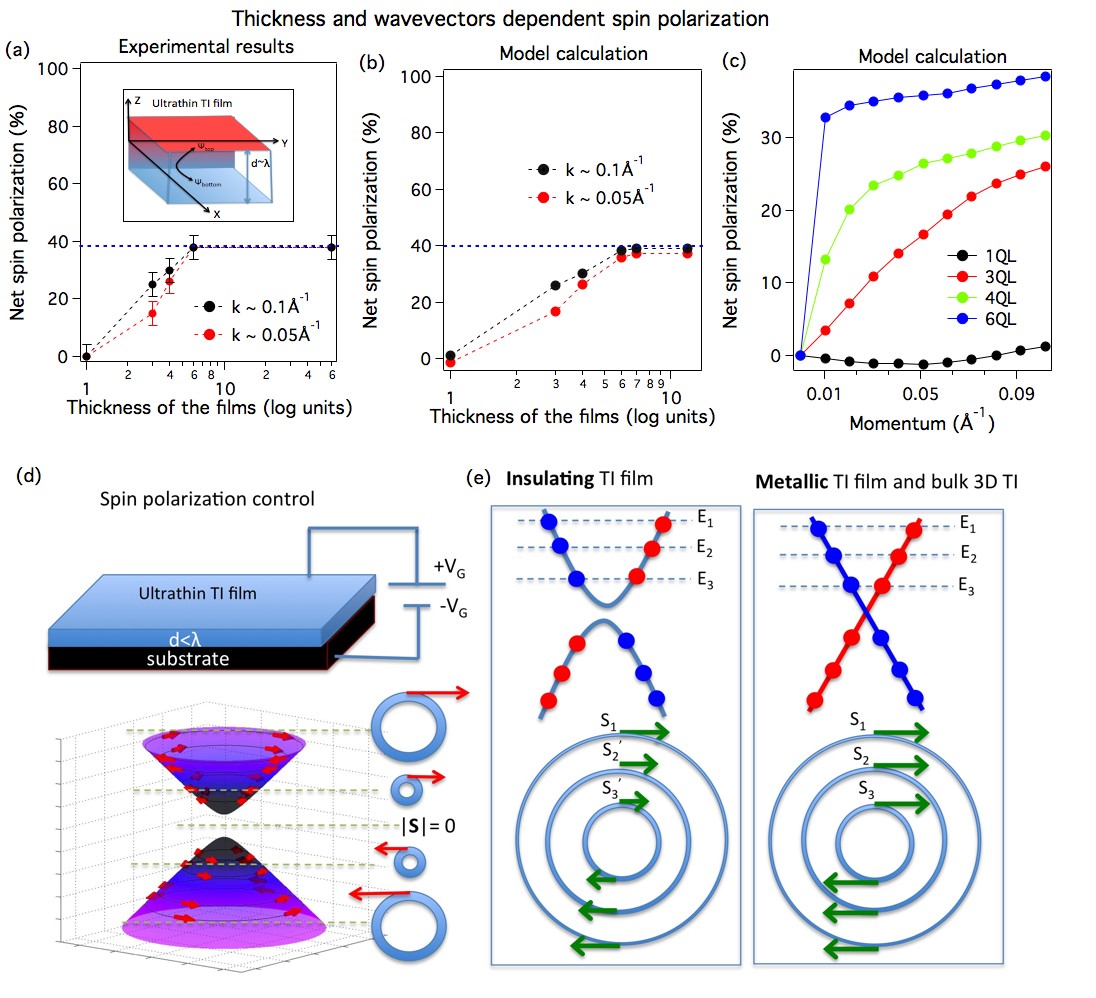}
\caption{\textbf{ Experimental versus theoretical spin polarization and texture.}
(a) Experimentally-observed net spin polarization as a function of thin film thickness for $k \sim$ 0.05  \AA$^{-1}$ and 0.1 \AA$^{-1}$. The inset shows a schematic view of quantum tunneling in ultrathin TI  films.
(b) Calculation results of spin polarization versus film thickness at two momentum points $k \sim$ 0.05  \AA$^{-1}$ and 0.1 \AA$^{-1}$.
(c) Results of a calculation of the net spin polarization as a function of wavevector for ultrathin films of thickness 1QL, 3QL, 4QL, and 6QL.
Dashed lines in (a,b) and solid lines in (c) between the dots serve as guides to the eye. Error bars in (a) represent experimental uncertainties in determining the spin polarization. Schematic of  (d) a gate controlled spin polarization current switch device and
(e) momentum dependent spin configuration in ultrathin (insulating) film and thicker (metallic) film. }
\end{figure*}

\setcounter{figure}{0}

\renewcommand{\figurename}{\textbf{Supplementary figure}}

\clearpage

\textbf{
\begin{center}
{\Large \underline{Supplementary Information}: \\
Observation of Quantum-Tunneling Modulated Spin Texture in Ultrathin Topological Insulator Bi$_2$Se$_3$ Films}
\end{center}
}

\vspace{0.2cm}


\vspace{0.25cm}


\textbf{This files includes:}

\textbf{
\begin{tabular}{l l}
  Supplementary Figures \\ 
   Supplementary Notes \\
  Supplementary Methods\\
 Supplementary References \\
\end{tabular}}

\clearpage

\subsection{\large {Supplementary Figures}}


\begin{figure*}[h]
\centering
\includegraphics[width=15cm]{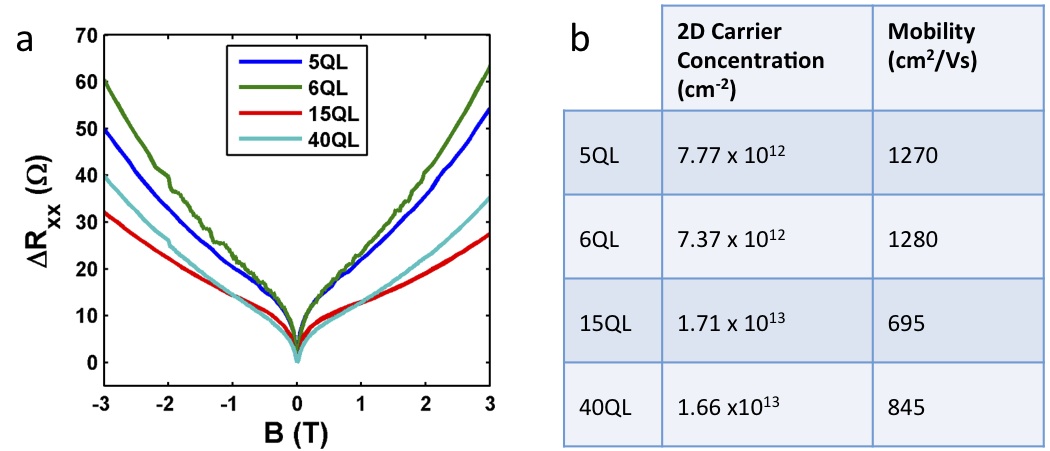}
\caption{(a) Magnetotransport results taken from thin Bi$_2$Se$_3$ films with various thicknesses shown in QL units. (b) Carrier concentration and mobility for various thicknesses.}
\end{figure*}

\clearpage

\begin{figure*}[h]
\centering
\includegraphics[width=11.50cm]{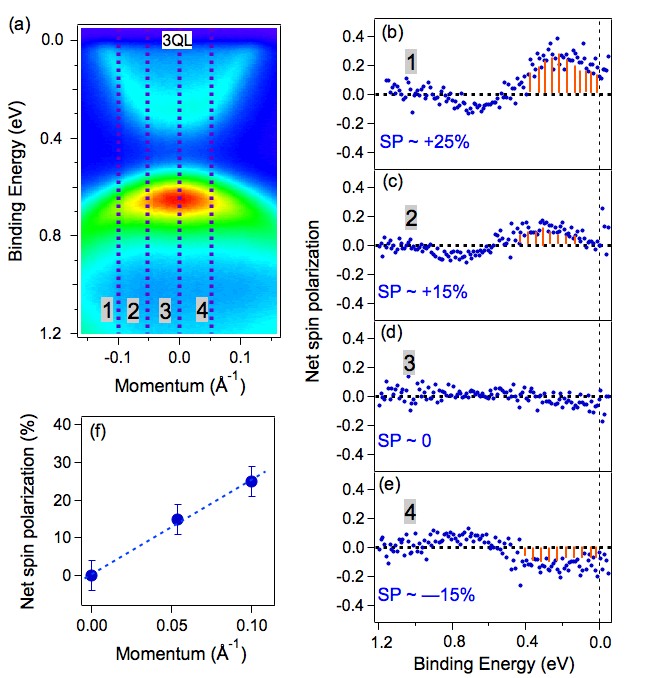}
\caption{Momentum dependent spin configuration.
(a) High-resolution ARPES spectrum of a representative 3QL Bi$_2$Se$_3$ film near the $\bar{\Gamma}$ point along the $\bar{\Gamma}-\bar{\text{K}}$ cut-direction. The blue dashed lines with integer numbers mark wavevectors at which SR-ARPES measurements were done. 
(b-e) show the 
total tangential spin polarization for the wavevectors marked in  (a) where blue curve shows the polarization magnitude as a function of binding energy. 
The value of net spin polarization (SP) is noted on the plot. The vertical red lines are guides to the eye, indicating a non-zero area under the spin polarization curve. (f) The net tangential spin polarization magnitude versus wavevector, 
which suggests that the spin configuration of the material strongly depends on wavevector in the ultrathin limit.
The blue dashed line serves as a guide to the eye. Error bars represent experimental uncertainties in determining the spin polarization.
}
\end{figure*}

\clearpage

\begin{figure*}[h]
\centering
\includegraphics[width=16cm]{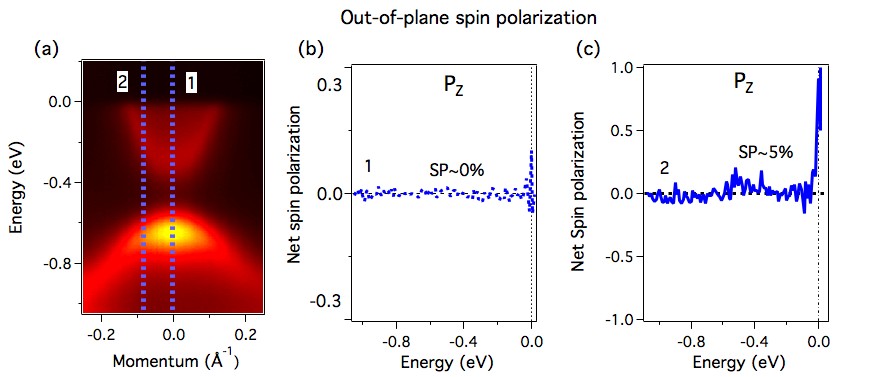}
\caption{Out-of-plane spin polarization in a 3QL Bi$_2$Se$_3$ film. (a) Spin-integrated ARPES spectrum of surface states. The blue dashed lines and number indicate the momenta position at which SR-ARPES measurements were carried out. (b) Out-of-plane spin polarization at momentum position $\#1$. (c) Same as (b) for momentum position $\#2$. Negligible out-of-plane spin polarization is observed.
These measurements were performed in I3 beamline at Maxlab.}
\end{figure*}

\clearpage

\begin{figure*}[h]
\centering
\includegraphics[width=15cm]{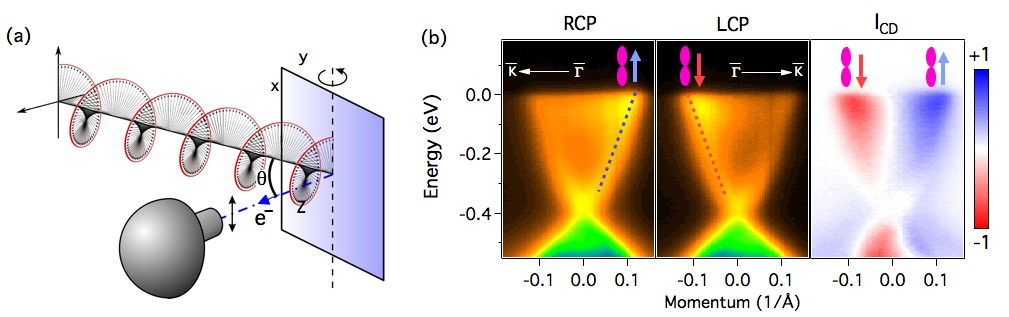}
\caption{CD-ARPES measurements. (a) Experimental geometry for circular dichroism (CD) measurements. (b) CD data for bulk Bi$_2$Se$_3$ measured at 19 eV. The red/blue arrow sign shown on the spectra represent the direction of the spin (down/up) at particular momentum. Similarly, the dumbbell signs on the spectra indicate that the current experimental geometry mainly probes the $p_z$ orbitals of Bi and Se.}
\end{figure*}

\clearpage

\begin{figure*}[h]
\centering
\includegraphics[width=16cm]{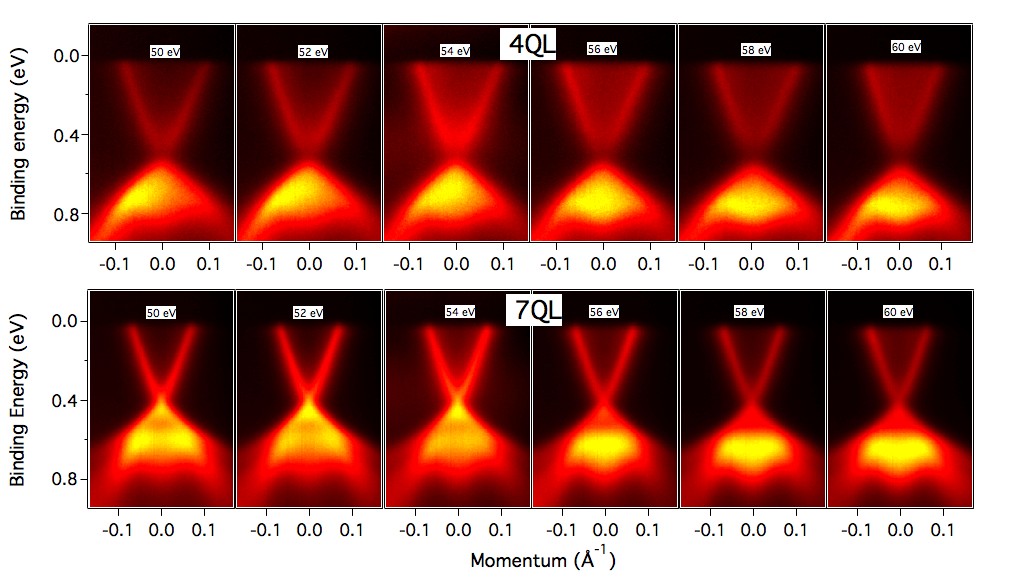}
\caption{Photon energy dependent ARPES spectra of ultrathin 4QL (top row) and 7QL (bottom row) Bi$_2$Se$_3$ films measured  along the
$\bar{\Gamma}-\bar{\text{M}}$ momentum cut at temperature T = 20 K. The measured photon energies are noted on the plots.}
\end{figure*}

\clearpage

\begin{figure*}[h]
\centering
\includegraphics[width=17cm]{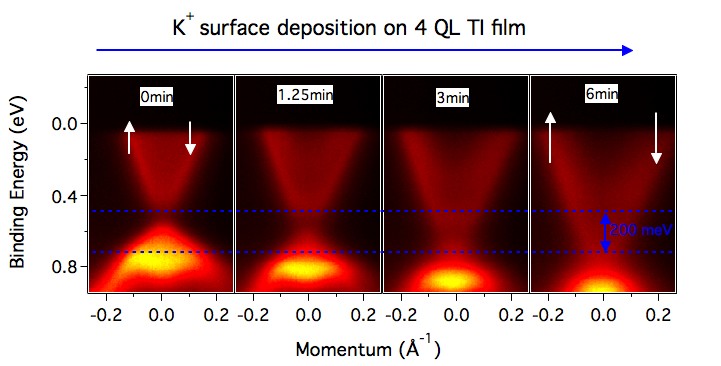}
\caption{ARPES spectra of a 4QL Bi$_2$Se$_3$ film along the
$\bar{\Gamma}-\bar{\text{K}}$ momentum cut for different potassium deposition times. The time evolution of potassium deposition is indicated at the top of each panel. The deposition rate is approximately 0.1 \AA min$^{-1}$. Spectra were measured with a photon energy of 60 eV at a temperature of 20 K.}
\end{figure*}

\clearpage

\begin{figure*}[h]
\centering
\includegraphics[width=17cm]{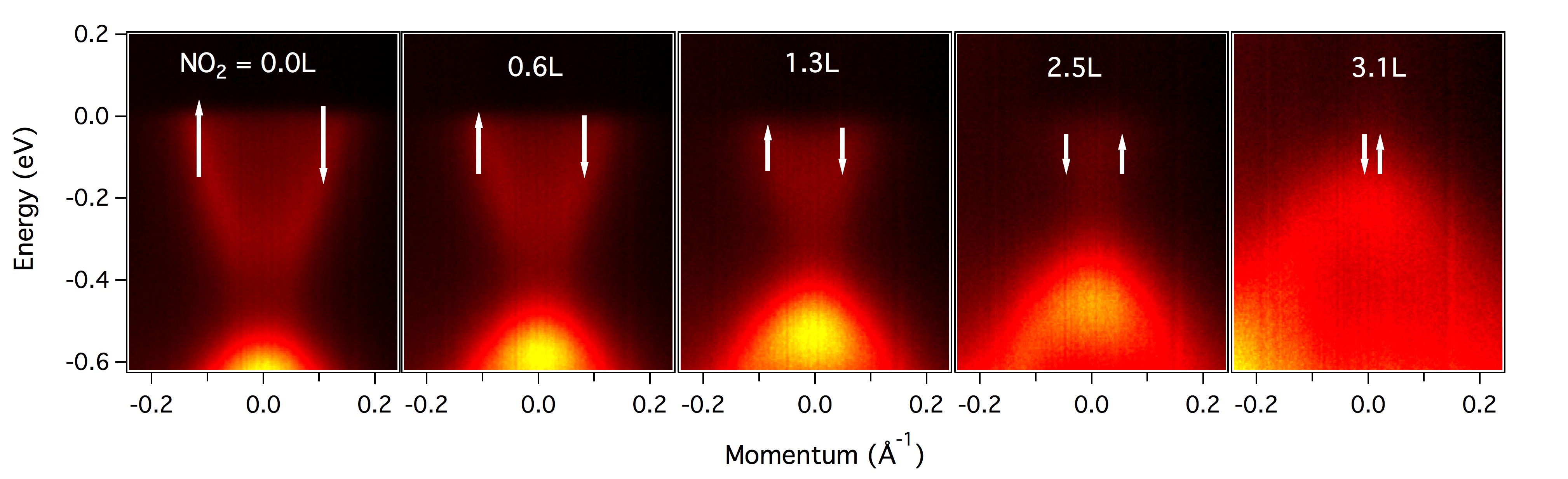}
\caption{Chemical potential tuned to lie inside the tunneling gap. Measured surface state dispersion on \textit{in situ} NO$_2$ surface adsorption on the
Bi$_2$Se$_3$ film surface. The NO$_2$ dosage is in Langmuir units (1.0 L=6$\times$$^{-6}$ torr s).
ARPES spectra of a 3QL Bi$_2$Se$_3$ film along the
$\bar{\Gamma}-\bar{\text{K}}$ momentum cut for different NO$_2$ dosages. The white arrows represent the magnitude of the spin polarization, which decrease with increasing NO$_2$ adsorption. Spectra were measured with a photon energy of 55 eV at a temperature of 20 K.  }
\end{figure*}

\clearpage


\subsection{\large {Supplementary Note 1}}
\vspace{0.1cm}

\textbf{Transport measurements of thin films}

We performed systematic magnetotransport measurements for TI thin films with various thicknesses. 
Films were patterned into Hall bars using photolithography and measured using a standard AC lock-in at liquid helium temperatures. The change in magnetoresisistance for different Bi$_2$Se$_3$ film thicknesses is shown in Supplementary Figure 1. An increasing weak-antilocalization component of the total magnetoresisistance is seen as the films become thinner. This correction is described by the Hikami-Larkin-Nagaoka equation and is due to conduction through the surface states. Fits yield values of alpha between -0.5 and -1 and phase breaking lengths of approximately 100 to 200 nm. The 2D carrier concentration was calculated from linear fits to the Hall voltage. It rises for thicker films but does not scale linearly the thickness, indicating a large component of the carriers are in the 2D surface state. The results are consistent with published results [1, 2].


\subsection{\large {Supplementary Note 2}}



\vspace{0.1cm}

\textbf{ Spin polarization measurements of ultrathin 3QL film}

\vspace{0.1cm}

We investigate the degree of spin polarization of the films as a function of electron wavevector.
Based on these data presented in Supplementary Figure 2, it is evident that the spin signal is dominated by the contribution from the  top surface since it reveals a characteristic left-handed helical spin-texture (see the spin polarization at locations \# 2,4) signaling topological order  [3]. For smaller wavevectors, net spin polarization is seen to decrease (locations \#2,4 in Supplementary Figure 2(a)).
The decrease can be understood as the presence of a tunneling gap in the ultrathin limit that effectively prevents the partner-switching behavior expected in the gapless topological surface states system [4]. The tunneling gap for ultrathin films can be seen in  the data (Supplementary 
Figure 2(a)).  The reduction of spin polarization and the existence of a tunneling gap in the data suggest that the bottom surface contribution of spin polarization must increase to effectively cancel the left-handed helical spin upon approaching smaller momenta values toward $\bar{\Gamma}$.
Similarly at the time-reversal invariant point (point \#3 in Supplementary Figure 2(a)), namely at $\bar{\Gamma}$, a nearly ideal cancellation between top and bottom surfaces happens  resulting in zero net spin polarization, which is consistent with the time-reversal symmetry of the electronic structure of the films [4].

\vspace{0.3cm}

\textbf{Absence of the out-of-plane spin component in ultrathin TI film}

\vspace{0.3cm}

Here we present an out-of-plane spin measurement of a 3QL ultrathin TI film performed at the I3 beamline at Maxlab [5]. 
Supplementary Figure 3(a) shows the spin-integrated ARPES data with blue dashed lines and integer numbers indicating the momentum positions for the spin-resolved measurements. The net spin polarization is shown in Supplementary Figure 3(b) and 3(c) for momenta positions mark as $\#$ 1 and 2, respectively. Our spin polarization result in Supplementary Figure 3(b) indicates that the out-of-plane spin polarization is zero at the time-reversal invariant point ($\bar{\Gamma}$), which is in sharp contrast to magnetically doped topological insulator [6]. The absence of an out-of-plane spin component at $\bar{\Gamma}$ implies that the origin of gap is purely quantum tunneling between the top and bottom surfaces of the ultrathin TI. Similarly, a small out-of-plane spin component is observed at momentum position $\#$ 2 , which is within the experimental resolution.
The negligible out-of-plane component of spin polarization at momentum position away from the time-reversal symmetry breaking point (Supplementary Figure 3(c)) indicates near absence of the Fermi surface warping on the 3QL film Dirac cone.

\vspace{0.3cm}



\vspace{0.1cm}

\textbf{CD-ARPES measurements}

The experimental geometry used for the circular dichroism (CD)  measurements is present in Supplementary Figure 4(a). The sample surface is parallel to the XY plane and circularly polarized photons (spiral arrow) propagate in the XZ plane at an angle ($\theta$) of 50$^{\circ}$ to the sample surface normal. The ARPES dispersion maps of surface states measured using right circularly polarized (RCP) light and left circularly polarized (LCP) light for bulk Bi$_2$Se$_3$ are shown in Supplementary Figure 4(b).
A clear surface state CD response on the photoelectron signal from the upper Dirac cone is observed where the +k Dirac branch shows stronger response for RCP light and the -k Dirac branch shows  stronger response for LCP light (Supplementary Figure 4(b)). 
The magnitude of the CD response signal defined as $I_{CD}$=($I_{RCP}-I_{LCP}$)/($I_{RCP}+I_{LCP}$) is observed to be about 30\% for incident photons with an energy of 19 eV in Bi$_2$Se$_3$ for electrons with binding energy of about 100meV, well below the chemical potential. 
This CD behavior is qualitatively consistent with previous work [7, 8]. 
 Even though the p$-$polarized light probes both in-plane ($p_x$, $p_y$) and out-of plane ($p_z$) orbitals revealing groundstate of the topological states, the topological surface states has dominant $p_z$ orbital character (up to 80\%) [9, 10]. 
Based on our experimental geometry and the light polarization, the observation of strong CD response in bulk Bi$_2$Se$_3$ is likely coming from the $p_z$ orbitals of Bi and Se.

\vspace{0.3cm}

\subsection{\large {Supplementary Note 3}}

\textbf{Chemical potential tuning in ultrathin TI films}

It is important to note that 
a wide range of electronic structures have been reported in ultrathin TI films grown by MBE depending on the nature of the substrates used  [11-13]. 
Different substrates result in different potential gradient from the substrate (the bottom surface of the ultra-thin film) to the vacuum (the top surface of the film). When the potential is large, the Dirac point energy of the bottom surface is offset with respect to that of the top surface. Such Dirac point energy offset is observed to cause sizable Rashba-type splitting of surface states as reported in ultrathin Bi$_2$Se$_3$ films grown on
double-layer-graphene-terminated 6H-SiC (0001) substrate [11] and Bi-terminated Si(111)-(7$\times$7) [13] substrate by MBE, respectively. On the other hand, no observable Rashba-type splitting due to substrate potential is reported in the Bi$_2$Se$_3$ grown on Si(111)$\beta\sqrt{3}\times\sqrt{3}$-Bi substrate (in ref [12]).

The photon energy dependent ARPES spectra for 4QL and 7QL films  of Bi$_2$Se$_3$ are presented in Supplementary Figure 5. Nearly parabolic electronic band structure is observed in the binding energy of 500 meV for all photon energies with no observable additional features such as Rashba-like splitting in the 4QL film. Similarly for 7QL Bi$_2$Se$_3$ film which is just above the quantum tunneling limit, the Dirac like surface states do not disperse with varying photon energy. These photon energy dependent studies suggest that our GaAs(111)A substrate is similar to Si(111)$\beta\sqrt{3}\times\sqrt{3}$-Bi substrate in ref [12], which implies nearly absence of the substrate-induced potential difference between the two surfaces.
Nonetheless, we emphasize that for possible device applications it is essential to understand how the substrate-induced potential modifies the surface state dispersion relation at the interface between the substrate and the TI.


As discussed in the main text, our systematic results suggest that by varying the gate voltage, the spin polarization can be tuned to zero. This effect, which cannot be readily achieved using the highly-polarized surface states of bulk TIs, allows ultrathin film TIs to serve as the basis for a spintronics device which switches between a spin-polarized and non-polarized state depending on the gate voltage. This phenomena emphasizes the importance of understanding how spin configuration in topological materials is coupled to effects such as quantum tunneling. 

Here we demonstrate an example of surface gating by \textit{in situ} surface chemical treatments such as surface deposition. Supplementary Figures 6 and 7 present the energy dispersion, measured by ARPES, along the $\bar{\Gamma}-\bar{\text{K}}$ momentum cut during potassium deposition and NO$_2$ adsorption on a 4QL and 3QL ultrathin Bi$_2$Se$_3$ film, respectively. The potassium deposition surface treatment acts to dope the surface with electrons, raising the surface chemical potential by nearly 200 meV in 4QL film. This drives the system to a state of high surface spin polarization magnitude. On the other hand, NO$_2$ adsorption moves the chemical potential to in-gap region, lowering the surface chemical potential by nearly 300 meV in 3QL film.

Electrostatic gating with a top or back gate wouldn't be significantly different than the in-situ chemical doping as shown in Supplementary Figures 6 and 7, where the surface state structure didn't change appreciably, while the Fermi energy is moved from regions of high polarization to low polarization. However, if there was an appreciable change in the coupling between the two surfaces this could be overcome by simultaneously gating the top and back surfaces to move the potential of both surfaces by the same amount as shown in ref [14].  Regardless of the technical details of proposed thin film devices, the energy-momentum space spin texture revealed in our study provides a critical knowledge to design and interpret the functional devices based on ultrathin films.

Furthermore, it is important to note that successive deposition of potassium on the surface of ultrathin Bi$_2$Se$_3$ film makes the ARPES spectrum broader.
In spite of the broadened spectrum due to K$^+$ ion coverage on the surface, it still can be observed that the gap at the Dirac point becomes smaller. While it is reasonable to assume that the electrons brought by potassium go into the film relatively homogeneously in $\hat{z}$ axis, K$^+$ ions stay on the surface and thus act as an effective gating that changes the potential difference between the top and the bottom surfaces. 
Similar broadening of the spectra also happened to be observed during NO$_2$ adsorption.



\subsection{{\large Supplementary Methods }}

\textbf{Surface deposition:} Potassium deposition was performed at beamline 12.0.1 of the ALS from a SAES getter source
(SAES Getters USA, Inc.), which was thoroughly degassed before the experiment. Pressure in the experimental chamber stayed below $1\times 10^{-10}$ torr during deposition. The deposition rate ($\mathrm{\AA}$/sec) was monitored using a commercial quartz thickness monitor (Leybold Inficon Inc., Model XTM/2). The deposition amount (thickness) was obtained by multiplying the deposition rate by the elapsed time.

Adsorption of NO$_2$ molecules on surface of ultrathin 3QL Bi$_2$Se$_3$ film was achieved via controlled exposures to NO$_2$ gas (Matheson, 99.5$\%$) at beamline 12.0.1 of the ALS. The adsorption effects were studied under static flow mode by exposing the clean sample surface to the gas for a certain time at the pressure of $1\times 10^{-8}$ torr, ARPES spectra were measured after the chamber was pumped down to the base pressure. Spectra of the NO$_2$ adsorbed surfaces were taken within minutes of opening the photon shutter to
minimize potential photon induced charge transfer and desorption effects.


\textbf{CD-ARPES measurement:} For CD-ARPES measurements, single crystalline samples of topological insulators were grown using the Bridgman method, which is detailed elsewhere [15]. ARPES measurements for the low energy electronic structure were performed at the Synchrotron Radiation Center (SRC), Wisconsin, equipped with high efficiency VG-Scienta SES2002 electron analyzer.
The polarization purity is better than 99\% for horizontal polarization (HP) and better than 80\% for RCP and LCP.
Samples were cleaved {\it in situ} and measured at 20 K in a vacuum better than 1 $\times$ 10$^{-10}$ torr.
Energy and momentum resolution were better than 15 meV and 1\% of the surface Brillouin zone (BZ), respectively. 

\textbf{ Theoretical calculation methods:}
Theoretical calculations presented in the main text are based on the generalized
gradient approximation (GGA) [16] using the full-potential projected
augmented wave method [17, 18] as implemented in the VASP package [19]. We use 1QL, 3QL, 4QL, 6QL, 7QL, and 12QL slab models with a
vacuum thickness larger than $10 \mathrm{\AA}$. The electronic structure calculations were performed over a $11 \times 11 \times 1$ Monkhorst-Pack $k$-mesh with the spin-orbit coupling included self-consistently.

In order to calculate the spin polarization, we consider the spin expectation value 
$\langle{S_\alpha}_(i,k)\rangle ={\sum_\tau}\langle\psi_{i,k,\tau}|\sigma_\alpha|\psi_{i,k,\tau}\rangle $, where $i$ = atomic layer index, $k$ = momentum vector, $\tau$ = orbital index, $\sigma$ = Pauli matrix, and $\alpha$ = x, y, z. 
To simplify the calculation, (1) we choose momentum vector k along high symmetry direction $k_x$. Due to the mirror symmetry, we have$\langle {S_x} \rangle_{i,k} =0$, 
and (2) only focus near Dirac point where the hexagonal warping effect is small, so $\langle{S_z}\rangle_{i,k}$  can be neglected. Therefore, only y components survive. 
We ignore subindex y to simplify notation, $\langle {S_\alpha}\rangle_{i,k} =\langle S\rangle_{i,k}$. In the calculation we also consider the electron attenuation length ($\lambda$) due to the scattering processes since only electrons near the surface are able to reach the top of the surface and escape into the vacuum as in the measurement condition. 
Indeed, the spin polarization obtained by ARPES reflects the spin-texture of the states associated with top surface rather than the bottom surface. The calculated spin polarization for the electrons that can escape from the sample is $\langle S\rangle_k ={\sum_i}\langle S\rangle_{i,k} exp⁡(-{d_i}/\lambda)$ with proper normalization, 
where $d_i$  is the distance of an atom to the top surface, and the $\langle S\rangle_{i,k}$  is the spin expectation value for each atom and k point. The contribution from each atom is weighted by $exp⁡(-{d_i}/\lambda)$, which reduces the contribution from the bottom layer. Figs. 4(b) and 4(c) show the calculated results with $\lambda= 8\AA$.


\subsection{\large {Supplementary References}}

\begin{tabular}{l l}

$[1]$& Richardella, A. \textit{et al.} Coherent heteroepitaxy of Bi$_2$Se$_3$ on GaAs (111)B.\\& \textit{Appl. Phys. Lett.} $\mathbf{97}$, 262104 (2010).\\

 $[2]$& Taskin, A. A. \textit{et al.} Manifestation of topological protection in transport\\& properties of epitaxial Bi$_2$Se$_3$ thin films. \textit{Phys. Rev. Lett.} $\mathbf{109}$, 066803 (2012).\\

$[3]$&Hsieh, D. \textit{et al}. A tunable topological insulator in the spin helical Dirac transport\\& regime. \textit{Nature} $\mathbf{460}$, 27 (2009).\\

$[4]$& Lu, H.-Z. \textit{et al}. Massive Dirac fermions and spin physics in an ultrathin film of \\& topological insulator. \textit{Phys. Rev. B} $\mathbf{81}$, 115407 (2010).\\

$[5]$& Berntsen, M. H. A spin- and angle-resolving photoelectron spectrometer.\\&  \textit{Rev. Sci. Instrum} $\mathbf{81}$, 035104 (2010).\\

$[6]$& Xu, S.-Y. \textit{et al}. Hedgehog spin texture and BerryÕs phase tuning in a magnetic\\& topological insulator.   \textit{Nature Phys.} $\mathbf{8}$, 616-622 (2012).\\

$[7]$& Neupane, M. \textit{et al.} Oscillatory surface dichroism of the insulating topological\\& insulator Bi$_2$Te$_2$Se. \textit{Phys. Rev. B} $\mathbf{88}$, 165129 (2013).\\

$[8]$& Wang, Y. and Gedik, N. Circular dichroism in angle-resolved photoemission\\& spectroscopy of topological insulator. \textit{Phys. Status Solidi RRL} $\mathbf{7}$, 64 (2013).\\

 $[9]$& Zhu, -H. Z. \textit{et al.} Layer-by-layer entangled spin-orbital texture of the\\& topological surface state in Bi$_2$Se$_3$. \textit{Phys. Rev. Lett.} \textbf{110}, 216401 (2013).\\

$[10]$& Park, S. R. \textit{et al.} Chiral orbital-angular momentum in the surface states\\& of Bi$_2$Se$_3$. \textit{Phys. Rev. Lett.} $\mathbf{108}$, 046805 (2012).\\

$[11]$& Zhang, Y. \textit{et al}. Crossover of the three-dimensional topological insulator\\& Bi$_2$Se$_3$ to the two-dimensional limit.  \textit{Nature Phys.} $\mathbf{6}$, 584-588 (2010).\\

$[12]$&Sakamoto, Y. \textit{et al}. Spectroscopic evidence of a topological quantum phase\\& transition in ultrathin Bi$_2$Se$_3$ films. \textit{Phys. Rev. B} $\mathbf{81}$, 165432 (2010).\\

$[13]$& Berntsen, M. H. \textit{et al.} Direct observation of decoupled Dirac states at the interface \\&between  topological and normal insulators. Phys. Rev. B $\mathbf{88}$, 195132 (2013).\\

$[14]$& Liu, H. \&  Ye, P. D.  \textit{et al.}, Atomic-layer-deposited Al$_2$O$_3$ on Bi$_2$Te$_3$ for topological \\& insulator field-effect transistors.  \textit{App. Phys. Lett.} $\mathbf{99}$, 052108 (2011).\\

$[15]$& Jia, S. \textit{et al.}, Defects and high bulk resistivities in the Bi-rich tetradymite \\&topological insulator Bi$_{2+x}$Te$_{2-x}$Se. \textit{Phys. Rev. B} $\mathbf{86}$, 165119 (2012).\\

 \end{tabular}

\begin{tabular}{l l}



 
$[16]$&Perdew, J. P., Burke, K. \& Ernzerhof, M. Generalized gradient approximation\\&  made simple.  \textit{Phys. Rev. Lett.} \textbf{77}, 3865 (1996).\\

$[17]$& Bl$\ddot{o}$chl, P. E. Projector augmented-wave method. \textit{Phys. Rev. B} \textbf{50}, 17953 (1994).\\

$[18]$&G. Kresse, G. \& Joubert, J.  From ultrasoft pseudopotentials to the projector\\& augmented-wave method.  \textit{Phys. Rev. B} \textbf{59}, 1758 (1999).\\

$[19]$& Kress, G. \& Hafner, J. Ab initio molecular dynamics for open-shell transition\\& metals.  \textit{Phys. Rev. B} \textbf{48}, 13115 (1993).\\

 \end{tabular}

\vspace{0.8cm}
Correspondence and requests for materials should be addressed to M.Z.H. (Email:
mzhasan@princeton.edu).

\end{document}